\documentclass[journal]{IEEEtran}

\ifCLASSINFOpdf
\else
\fi
\usepackage{amssymb}
\usepackage[dvips]{graphicx}
\usepackage{multirow}
\usepackage{algorithm}
\usepackage{algorithmic}

\newcommand{\tabincell}[2]{\begin{tabular}{@{}#1@{}}#2\end{tabular}}

\begin{document}
%
\title{Envisioning Device-to-Device Communications in 6G}
%
%
%

\author{Shangwei~Zhang,~\IEEEmembership{Member,~IEEE,}
	Jiajia~Liu,~\IEEEmembership{Senior Member,~IEEE,}
	Hongzhi~Guo,~\IEEEmembership{Member,~IEEE,}
	Mingping~Qi,
	and~Nei~Kato,~\IEEEmembership{Fellow,~IEEE}
	\thanks{Shangwei Zhang, Jiajia Liu, Hongzhi Guo and Mingping Qi are with the School of Cybersecurity, Northwestern Polytechnical University. Corresponding author e-mail: liujiajia@nwpu.edu.cn.}
	\thanks{Nei Kato is with the Graduate School of Information Sciences, Tohoku University.}
}

\maketitle

\begin{abstract}
To fulfill the requirements of various emerging applications, the future sixth generation (6G) mobile network is expected to be an innately intelligent, highly dynamic, ultradense heterogeneous network that interconnects all things with extremely low-latency and high speed data transmission. It is believed that artificial intelligence (AI) will be the most innovative technique that can achieve intelligent automated network operations, management and maintenance in future complex 6G networks. Driven by AI techniques, device-to-device (D2D) communication will be one of the pieces of the 6G jigsaw puzzle. To construct an efficient implementation of intelligent D2D in future 6G, we outline a number of potential D2D solutions associating with 6G in terms of mobile edge computing, network slicing, and Non-orthogonal multiple access (NOMA) cognitive Networking.

\end{abstract}


%
\IEEEpeerreviewmaketitle

\section{Introduction}
%
%
%
%

The wireless communication system has been developed and regenerated for almost every decade to meet the ever-increasing network requirements. Until now, 5G has already been standardized in 2018 and its commercial deployment has been rolled out at the end of 2019. However, it is envisioned that 5G cannot fulfill the requirements of emerging Internet of Everything applications like augmented reality (AR), virtual reality (VR), mixed reality (MR), which require a convergence of communication, sensing, control, and computing functionalities \cite{Liu_CST2015}. In addition, new services, such as holographic communications, high-precision manufacturing, sustainable development and smart environments, enhanced energy efficiency, cannot be efficiently offered by current development of 5G networks \cite{Calvanese_VTM2019}. To fulfill such challenging requirements mentioned above, a novel network architecture of the future sixth generation (6G) mobile network is required. To date, debates on 6G have already been carried out in many countries and organizations. It might be too early to give an exact definition of 6G, nevertheless, most of researchers agree that 6G will be an innately intelligent, highly dynamic, extremely heterogeneous and ultradense network that interconnects all things. More concretely, 6G is expected to support data rate that is 100 to 1000 times faster than 5G (i.e., 1Tbps) with end-to-end latency less than 1 msec \cite{Letaief_CM2019}.

Based on the existing 5G architecture \cite{David_VTM2018}, current research trends have shown that 6G network architecture at least has the following characteristics: space-air-terrestrial-sea integrated (SATSI), artificial intelligence (AI) driven, higher frequency bands, ultradense heterogeneous, green energy, security and privacy. For the sake of ubiquitous connection, SATSI network architecture will enable flexible and efficient connection of trillion-level objects in terrestrial, aerial, space, and sea domains. To maintain network systems with ultra high data rate and extraordinarily low latency, 6G will utilize much broader frequency bands, including sub-terahertz and terahertz (THz) bands as well as visible light communication (VLC). Since higher frequency bands inevitably suffer from high path-loss, the data transmission distance will become much shorter. As a result, the 6G network will become extremely dense. Moreover, high-speed and low-latency D2D communication and ultra-massive MIMO communication will be the indispensable part in 6G nework to cope with the limited distance. Therefore, ultradense heterogenous network is considered to be one of the key components of 6G.

\begin{figure*}[!t]
	\centering
	\includegraphics[width=7in]{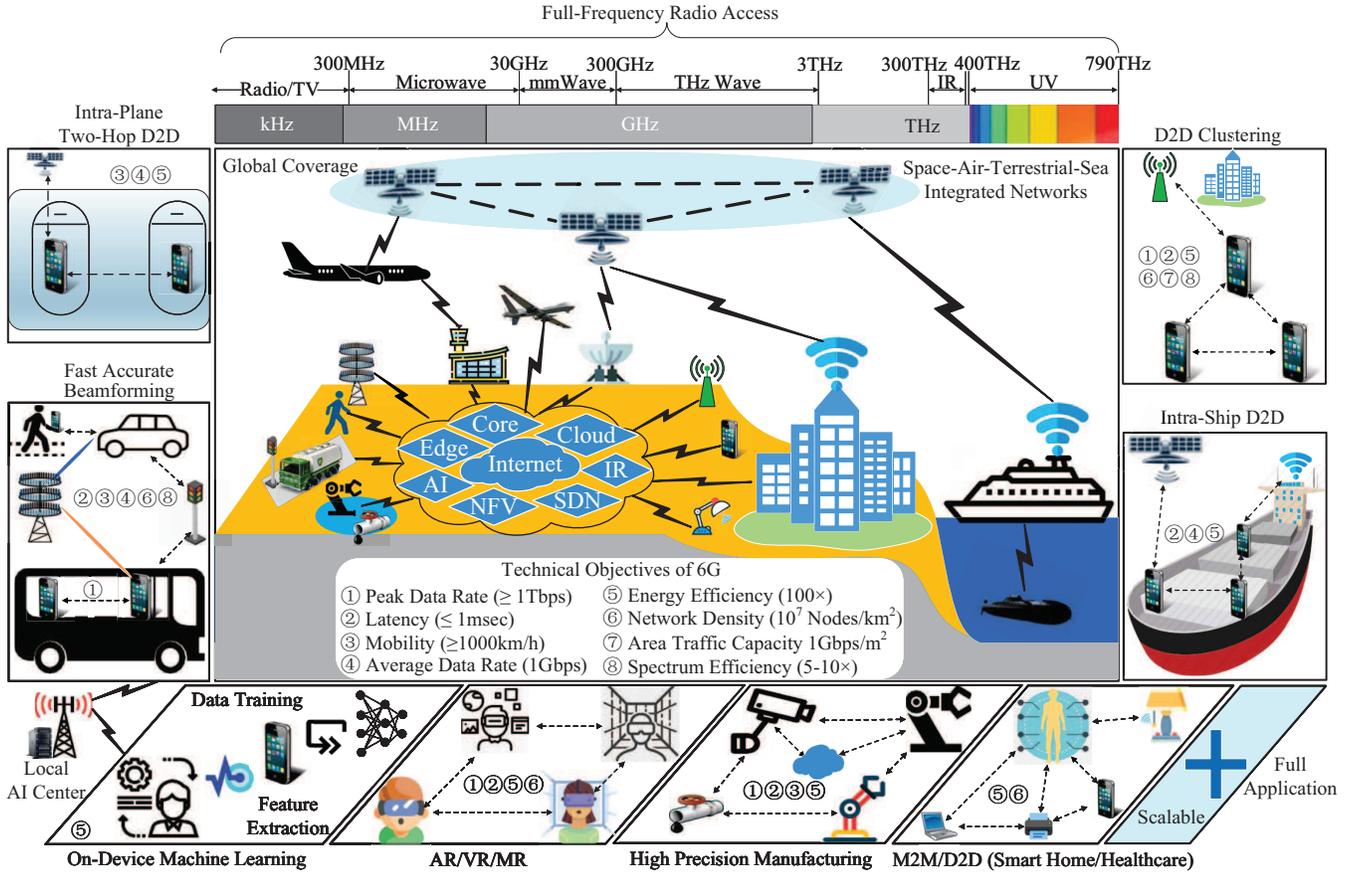}
	\caption{The D2D communications in future space-air-terrestrial-sea integrated 6G network.}
	\label{fig:6G_architecture}
\end{figure*}

Although 5G has benefited greatly from ultradense networks and device-to-device (D2D) communication, further network densification in 6G will encounter many serious problems and challenges \cite{David_VTM2019}, such as severe interference, very complex resource management, vast amount of signaling, prohibitively high cost and energy consumption, etc. Toward this end, researchers have pointed out that the utilization of AI techniques can help network operators to realize intelligent automated network operation, management and maintenance. It is believed that AI will be the most innovative technique for 6G networks \cite{Elsayed_VTM2019}. By sensing and network big data training, AI technique like machine learning (ML) will make complex 6G network more intelligent and can achieve high level of efficient network management and optimization, such as network environment sensing, status predicting, proactive configuring, dynamic optimizing, self-healing, etc. Besides, AI will gradually migrate from cloud to network edge, wherein multi-level AI will be employed\cite{Zong_VTM2019}.

It is noticed that mobile user equipments (UEs) like smartphones are becoming more and more powerful since they are integrated with powerful high-end processors, rich storage resources, long lifetime battery, multiple sensors, etc. They can be regarded as personal mobile workstations that enable light-level on-device AI processing for intelligent networking in future \cite{Letaief_CM2019}. Many researchers have pointed out that the utilization of on-device AI can dynamically sense local network environment, efficiently and optimally adjust network parameters to meet different network application requirements. Employing AI in D2D communication will lead to new opportunities and challenges in future 6G networks. As motivated by the aforementioned studies, we in this paper envision a number of potential intelligent D2D solutions associating with 6G in terms of intelligent mobile edge computing (MEC), network slicing as well as Non-orthogonal multiple access (NOMA) cognitive networking.

\section{Architectures and Developing Trend of 6G Networks}

\subsection{Space-Air-Terrestrial-Sea Integrated Network}
Note that relying only on current terrestrial network cannot fulfill the requirements for extremely broad coverage and ubiquitous connectivity. Therefore, 6G will further integrate sea/underwater networks with respect to 5G space-air-terrestrial network, so as to support various applications and maintain ubiquitous mobile ultrabroadband everywhere. As shown in Fig.~\ref{fig:6G_architecture}, a typical 6G SATSI network contains four parts: a space network composed of interconnected satellites with different orbits; an air network contains high altitude platform like airplane \cite{Huang_VTM2019}, air ship, as well as low altitude UAV and flying BSs; a heterogeneous ultradense terrestrial network that includes core network (CN), cloud, edge, and M2M communication networks; a sea-based network that contains shortwave communication, ultrashort-wave communication, submarine optical network and underwater acoustic network that provide wireless connections to those equipments over the sea and underwater area. It is believed that terrestrial network will still be the most important part for providing wireless mobile services for most human activities \cite{Zhang_VTM2019}.

The SATSI network can make full use of the characteristics of global network coverage, full-frequency radio transmission and full application to achieve seamless high-speed communication in space, air, terrestrial and sea domains, such as automated driving, high precision manufacturing, smart home/healthcare, AR, VR, MR, etc. With the participating of AI, D2D communication will be employed to support low latency and high speed data transmission in different network scenarios, such as intra-plane two-hop D2D, intra-ship multi-hop D2D, D2D with fast accurate beamforming in vehicular networks. In future 6G, massive D2D clusters will exist in local area due to the aggregation and social behavior of human beings. Users in each D2D cluster can be managed and served via D2D connection.

\subsection{Extremely Heterogeneous and Ultradense Terrestrial Network}
As wireless transmission distance will become much shorter in 6G, D2D communication is envisioned to continuously evolve into the 6G wireless communication systems to efficiently support a much larger and more diverse set of UEs and applications. Ultradense network integrated with D2D communication is considered as one of the pieces of the 6G jigsaw puzzle, making 6G terrestrial network inevitably be a ultradense heterogeneous network structure. Note that further network densification will lead to performance degradation due to the severe interference, prohibitively high cost and energy consumption \cite{David_VTM2019}. Besides, ultradense networks may lead to frequent handover for high speed moving UEs. As THz bands signal has very strong directional characteristics, the transmitter beam should perfectly point to the receiver antenna to successfully deliver the information, which makes the network management to be a very challenging task. Accordingly, emerging new interference management approaches, channel allocation scheme as well as spectral efficiency are still essential in 6G ultradense heterogeneous networks. Considering the frequent handover, interference and energy consumption problem in 6G ultradense network, D2D and NOMA based communication could be one feasible solution in future.

\subsection{Innately Intelligent and Highly Dynamic Network}
As 6G network will be much more complex and dynamic than the preceding generations of wireless networks, traditional network management methods will become untenable. Hence, intelligent network management and optimization approaches shall be employed to fulfill the different QoS demands. Over the past decade, many types of ML algorithms \cite{Elsayed_VTM2019}, like supervised learning, unsupervised learning, reinforcement learning, etc., were adopted for intelligent resource allocation, network optimization, sBSs or UAVs deployment, network design, etc. It is believed that AI-enabled radio access and core networks will allow future networks to sense and learn from environment, make decisions through big data training, automatically predict and adapt the network changes, and achieve optimal performance by self-configuration.

It is envisioned that multilevel distributed AI will be employed to provide global intelligent management of 6G networks. Specifically, global AI center will be deployed in core network, while local AI center can be embedded into traditional MBSs or MEC servers so as to provide AI proceeding on local network management. Noticed that UEs like smart phones, may provide opportunities for on-device data training, which can sense and learn from the local channel pattern, traffic pattern, moving trajectory, etc., to understand the features of user behavior and predict the network status. In future 6G, global AI, local AI and on-device AI will cooperate and leverage one another's advantages to maintain the network operation and optimization.

\section{Evolution of Terminal User Equipment in 6G}

\begin{figure}[!ht]
	\centering
	\includegraphics[width=3.5in]{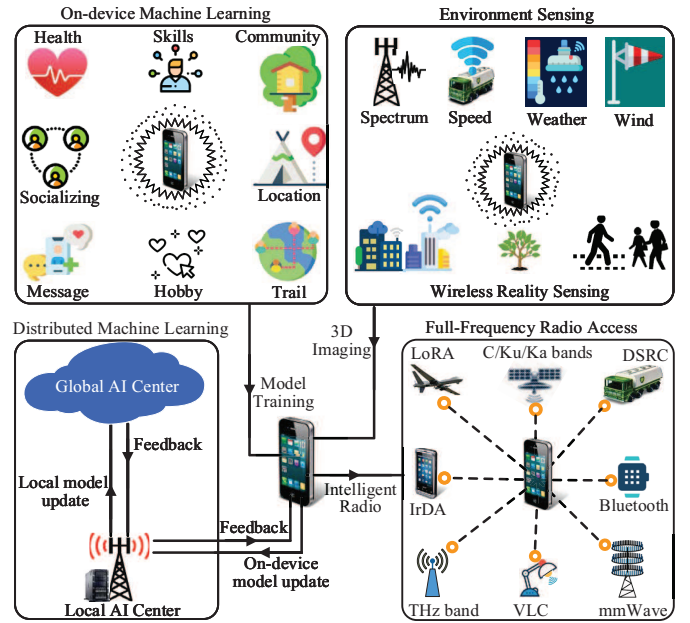}
	\caption{Illustration of the intelligent user equipment in 6G.}
	\label{fig:intelligent}
\end{figure}


\linespread{1.5}
\begin{table*}
	\centering
	\caption{Frequency resources available for D2D communications}
	\label{frequency_resource}
	\begin{tabular}{|c|c|c|c|c|c|c|}
		\hline
		\tabincell{c}{Access \\ Mode}&\tabincell{c}{System}&Frequency Band&\tabincell{c}{Network\\type}&\tabincell{c}{Transmission\\Range}&\tabincell{c}{Achievable\\ Rate}&\tabincell{c}{Application \\ scenarios} \\
		\hline
	\multirow{14}{*}{\tabincell{c}{Licensed\\Band}}	
		&2G&\tabincell{c}{890-960MHz}&GSM&35km&270.83kb/s&Macro Cell\\	
		\cline{2-7}		
		&3G&\tabincell{c}{1.94-2.145GHz}&WCDMA&10km&5.75-14.4Mb/s&Macro Cell\\		
		\cline{2-7}		
 		&4G&0.7-3.6GHz&LTE/LTE-A&1-5km&0.1-1Gb/s&Macro Cell\\	
 		\cline{2-7}
 		&5G&30-300GHz&mmWave&$<$1km&10Gb/s&\tabincell{c}{Massive MIMO, Macro/Pico Cell}\\	
 		\cline{2-7}
 		&6G&0.3-3THz&THz wave&$<$1km&1Tb/s&Small Cell\\			
 		\cline{2-7}		
 		&WMAN&2.3-3.6GHz&WiMax&50km&around 100Mb/s&Macro Cell\\		
 		\cline{2-7}
 		&IoT&\tabincell{c}{0.703-2.2GHz}&NB-IoT&$<$10km&160-250kb/s&M2M\\
 		\cline{2-7}	
		&\multirow{6}{*}{\tabincell{c}{Satellite}}
		&0.3-4GHz (UHF)&\tabincell{c}{L/S}&GEO/LEO&0.432-3Mb/s&\tabincell{c}{Space-air/ground/sea,\\Inter-Satellites, GNSS}\\
		\cline{3-7}
		&&4-30GHz (SHF)& \tabincell{c}{C/X/Ku/K/Ka}&\tabincell{c}{GEO/MEO/LEO}&1-200Mb/s&\tabincell{c}{Space-air/ground/sea, \\Feeder Link, Inter-Satellites}\\
		\cline{3-7}
		&&30-75GHz (EHF)& Ka/Q/U/V&MEO/LEO&several Gbps&\tabincell{c}{Space-air/ground/sea\\Feeder Link, Inter-Satellites}\\
		\cline{3-7}	  	
		\hline		
		\multirow{6}{*}{\tabincell{c}{Unlicensed\\Band}}		
		&\multirow{3}{*}{\tabincell{c}{WLAN/WPAN}}
		&2.4GHz,5GHz&WiFi&120m&11-600Mb/s&\tabincell{c}{Pico/Femto Cell, UAV}\\			
		\cline{3-7}
		&&2.4-2.4835GHz&Bluetooth&100m&around 1Mb/s&M2M\\
		\cline{3-7}
		&&430-790THz&Visible Light& $<$100m (LOS)&around 500Mb/s&Micro cell\\			
		\cline{2-7}		
		&\multirow{3}{*}{\tabincell{c}{IOT}}
		&0.125-13.56MHz&RFID/NFC&10cm&1-424kb/s&M2M\\
		\cline{3-7}
		&&300-400THz&IrDA&10m&around 4Mb/s&M2M\\
		\cline{3-7}			
		&&5.85-5.925GHz&DSRC (IoV)&1km&around 27Mb/s&\tabincell{c}{V2V, V2I, V2R, V2P}\\
		\cline{2-7}
		\hline	
	\end{tabular}
\end{table*}

\subsection{From Mobile Phone to Mobile Workstation}
Considering the growth of computational power provided by Moore's law and future 6G communication requirements, it can be seen that UEs will equipped with powerful high-end processors, extraordinary rich storage resources, plenty of sensors and ultralong lifetime batteries (ideally unlimited). When combined with breakthrough battery technologies and intelligent operating systems, the power of such UEs can be equivalent to personal mobile workstations. We envision that future UEs will be embedded with multiple radio interfaces with full frequency radio access capability that can connect to various network systems with different QoS requirements. Table~\ref{frequency_resource} summarizes the frequency resources available for D2D communications. One can see from Fig.~\ref{fig:intelligent} that future UEs can either connect to the nearby devices via short range communication technologies or establish long range links via satellite communications, thus ensuring the ubiquitous connection from anywhere at any time.

Equipped with plenty of sensors and powerful processors, future UEs have the real time collection capability for mapping and sensing the world, such as environment sensing, wireless reality sensing, etc. As higher frequency bands will be exploited in 6G, the antenna size on these spectrum bands can be so small that multiple antennas can be easily embedded in small size UEs or even in smaller IoT devices which may take advantages of the merits of MIMO technologies \cite{Zhou_WC2018}. Moreover, THz will enable new sensing applications in multiple domains like security sensing, gesture detection, body scanning, healthcare monitoring, as well as 3D mapping and imaging.

\subsection{From Smart mobile Phone to AI-driven mobile Phone}
The growing computation and storage power of devices can achieve on-device AI by local data training. According to \cite{Rappaport_Access2019}, smart phone will have computational capabilities that are on the order of the human brain in the near future. Due to the employment of higher frequency bands, UEs may equipped with multiple antennas and support full duplex transmission, high speed data transmission, large volume of data caching, MIMO communication, cognitive and sensing ability. All the above characteristics enable UEs to have AI ability to perform on-device ML so as to predict moving trace, common interest content, sensing local environment, etc. Therefore, mobile phone in 6G will shift from smart phone to AI-driven phone.

Accordingly, based on-device AI techniques, the future mobile phone will be designed not only for personal use, but also for intelligent networking, which can enable proactive 6G network management and automated network configuration, thus to improve personal network quality of experience (QoE) and quality of service (QoS). However, relying solely on on-device AI is far from enough to predict and manage the large domain dynamic networks. Local AI will be employed for local network data training, so as to intelligently utilize the collaboration among edge UEs and local AI center to exchange the learning parameters for a better training and inference of the models. In terms of data privacy and security, federated learning which can keep data training at each UE, will be employed to learn a shared global model from edge UEs.

\section{AI-Driven D2D Communication in 6G}
The key feature of D2D communication in 6G is to leverage the shared information from the local and on-device AI to extend UEs' ``view scope'' and increase their awareness of local network environment. To construct an efficient implementation of intelligent D2D in future 6G networks, we forecast that at least the following key techniques should be developed.

\begin{figure}[!h]
	\centering
	\includegraphics[width=3.5in]{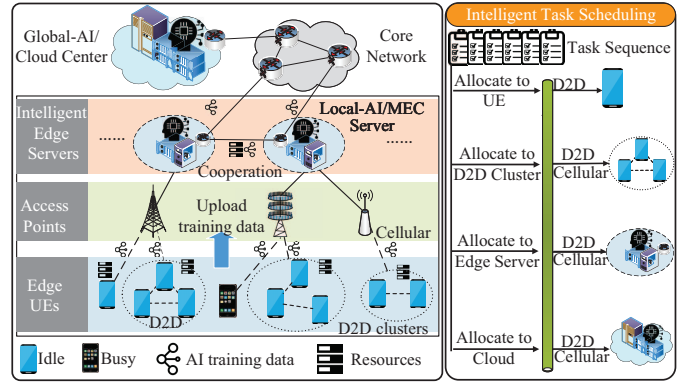}
	\caption{Illustration of the D2D-enhanced mobile edge computing.}
	\label{fig:edge_computing}
\end{figure}

\subsection{Intelligent D2D-Enhanced Mobile Edge Computing}
Alghouth 5G has benefited a lot from the joint cloud and MEC paradigm, however, edge servers may get congested or overloaded in 6G with plenty of emerging applications that require intensive computation, storage and transport resources. Moreover, the management and allocation of the resources will be a very difficult task. Noticed that the computing and storage resources of UEs are far from the requirement of users, the time for data transmission would be extremely short in 6G. Thus there would be a plenty of idle UEs in edge areas. Moreover, the utilization of THz band D2D communication would make the computing downloading and uploading between two close UEs near-real-time. Therefore, we envision a D2D-enhanced flexible approach that enables 6G local network to make use of the capacity of plenty idle UEs to enhance the network edge computing, as shown in Fig.~\ref{fig:edge_computing}.
Specifically, each UE in this architecture has a certain amount of computing and communication capabilities and can proceed task executions when being idle. Each task will be allocated to the idle UE first. If the capability of the single UE is relatively weak and could not be able to support the task, then it will further allocate the task to local D2D clusters via D2D or cellular links. If both the UE and D2D clusters cannot support the computation-intensive tasks, which will then be allocated to the edge servers or cloud servers that can offer more powerful computing capabilities \cite{Chen_Network2018}. The intelligent edge server combines local AI and MEC server together to intelligently provide services to the local edge UEs \cite{Zong_VTM2019}. The local AI and on-device AI can jointly manage the communication and computation resource, including sensing and predicting available dynamic D2D clusters' resources, optimizing and balancing the resource of the whole MEC systems, making decisions on task allocation.


\subsection{D2D-Enabled Intelligent Network Slicing}
In 5G, a network slice (NS) is a basic substrate offered by public land mobile network (PLMN) which is supplied by mobile network operators or infrastructure providers. Besides, network resources owned by private third-party actors outside of operator infrastructures can also be involved for network slicing \cite{Taleb_Network2019}. It is worth noticed that a large amount of dynamic D2D clusters can provide physical and/or virtual infrastructure resources, such as computation, network, memory, and storage, which may give rise to a proliferation of resources at the network edge. Therefore, we envision that global AI cooperating with local AI and on-device AI, will be used for real-time resource sensing and predicting to discover the dynamic resource surpluses provided by opportunistic D2D clusters, so that real-time dynamic network slicing will be proceed.

\begin{figure}[!ht]
	\centering
	\includegraphics[width=3.5in]{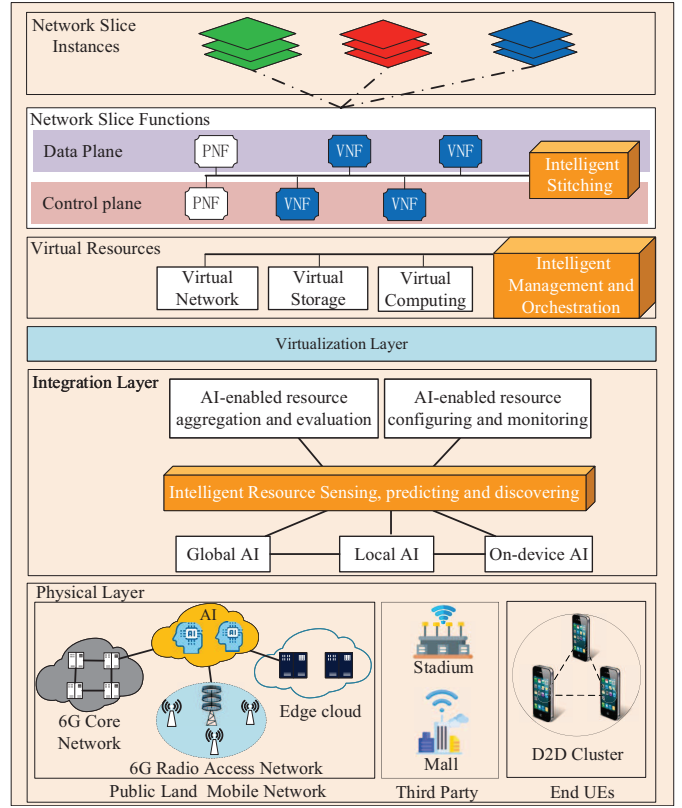}
	\caption{Illustration of D2D-enabled intelligent network slicing structure.}
	\label{fig:dynamic_slicing}
\end{figure}

As shown in Fig.~\ref{fig:dynamic_slicing}, the D2D enabled intelligent network slicing approach can help network operator to efficient federate and integrate network resources including PLMN, private third party and D2D clusters at network edge. To achieve intelligent network slicing, distributed AI will be employed in integration layer to predict and expose available dynamic network resources which are candidates to be integrated \cite{Thirdparty_NetSoft2018}. By merging the updates of AI models from UEs and exchanging the learning parameters, local AI can predict and provide real-time views of available dynamic resource information, including the available dynamic infrastructure resources, localization of available D2D clusters, etc. As infrastructure resources may come from different providers or stretch across greater geographical areas, AI techniques are employed for intelligent resource aggregating and evaluating, as well as configuring and monitoring. Thus, the required dynamic network resources can be identified by establishing the logical interconnection of available resources for operators. AI techniques are further employed to enable intelligent management and orchestration of virtual resources, as well as stitching physical network function (PNF) and virtual network function (VNF) to create the federated network slice instance.

\subsection{NOMA and D2D Based Cognitive Networking}
To support the exponentially increasing application demands and benefit from hybrid multiple access in future 6G, massive D2D clusters will be generated according to the aggregation and social behavior of human beings. UEs in each cluster can be intelligently managed and served via D2D connection by using NOMA, while different D2D clusters can employ orthogonal multiple access (OMA) for simultaneously communicating. When introducing NOMA into cognitive networks, the secondary user can efficiently cooperate with the primary user and obtain spectrum access opportunities to improve network spectrum efficiency and capacity. Engineers will encounter many challenges when employing NOMA in D2D, such as complex precoding mechanism in each NOMA group and co-channel interference management among UEs in the same subchannel. To this end, accurate beamforming based on intelligent channel estimation and dynamic precoding techniques will be employed for each UE to compensate for the additional power requirements that caused by the increased interference and noise.

\begin{figure}[!t]
	\centering
	\includegraphics[width=3.5in]{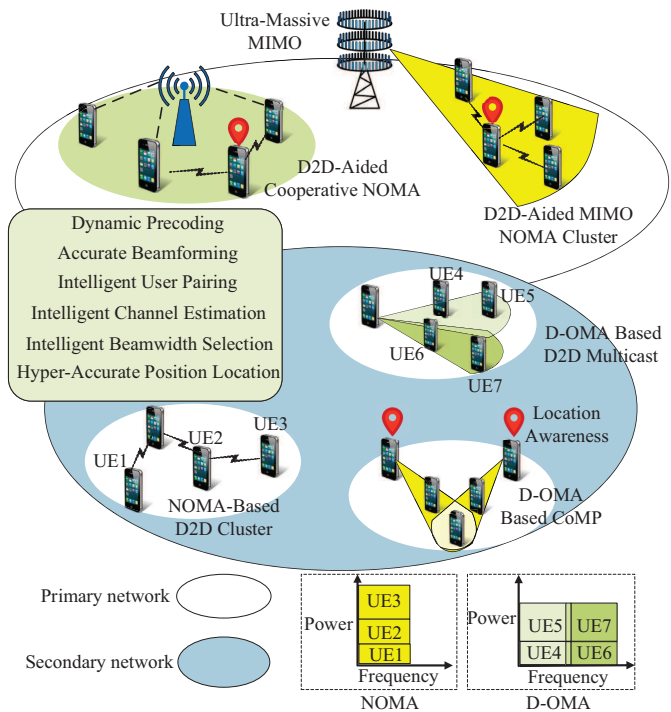}
	\caption{Illustration of the NOMA and D2D based cognitive networking.}
	\label{fig:d2d_noma}
\end{figure}

As shown in Fig.~\ref{fig:d2d_noma}, D2D-aided cooperative NOMA in primary networks can be achieved by selecting proper UE (resp. UEs) as NOMA relay (resp. relays) which can establish D2D connection to UEs in the nearby NOMA group. With the help of Local AI and on-device AI, the NOMA relays can be intelligently determined by utilizing techniques such as wireless cognition, intelligent user pairing, channel estimation, and hyper-accurate position location. Recently, the concept of delta-orthogonal multiple access (D-OMA) technique has been proposed for large scale concurrent access in future 6G cellular networks \cite{Al_VTM2019}. Although designed for cellular networks, D-OMA will be employed for D2D clusters in secondary networks to further improve network performance via accurate beamforming. The accurate beamforming can be achieved by utilizing multiple narrow beamwidth antennas or high frequency bands signal with strong directional characteristics. It can result in a highly dominant LOS component which is much stronger than the Non-LOS paths. The highly correlated channels at higher frequencies will make the use of D-OMA more attractive at higher frequencies.

\section{Conclusion}
We have envisioned in this article the D2D communications in future 6G. Based on the developing trend of potential 6G architecture and the evolution of future terminal user equipment, we focus our attention on conducting efficient implementation of intelligent D2D in future 6G communication systems, including D2D-enhanced mobile edge computing, D2D-enabled intelligent network slicing, as well as NOMA and D2D based cognitive networking. We hope our discussion will serve as guidelines for future D2D development in 6G.


%

%
%
%
%

\ifCLASSOPTIONcaptionsoff
  \newpage
\fi




\bibliographystyle{IEEEtran}
\bibliography{D2D_in_6G}

%

%
%
%






\end{document}